\documentclass[reprint,twocolumn, amssymb,aps,pra]{revtex4-1}

\usepackage{graphicx}
\usepackage{xcolor} 
\usepackage{amsmath}

\begin{document}
\title{Synthesizing Optical Spectra using Computer-Generated Holography Techniques}
\author{Connor M. Holland} 
\author{Yukai Lu}
\author{Lawrence W. Cheuk}
\email{lcheuk@princeton.edu}
\affiliation{Department of Physics, Princeton University, Princeton, NJ 08544 USA}
\date{\today}
\begin{abstract}
Experimental control and detection of atoms and molecules often rely on optical transitions between different electronic states. In many cases, substructure such as hyperfine or spin-rotation structure leads to the need for multiple optical frequencies spaced by MHz to GHz. The task of creating multiple optical frequencies -- optical spectral engineering -- becomes challenging when the number of frequencies becomes large, a situation that one could encounter in complex molecules and atoms in large magnetic fields. In this work, we present a novel method to synthesize arbitrary optical spectra by modulating a monochromatic light field with a time-dependent phase generated through computer-generated holography techniques. Our method is compatible with non-linear optical processes such as sum frequency generation and second harmonic generation. Additional requirements that arise from the finite lifetimes of excited states can also be satisfied in our approach. As a proof-of-principle demonstration, we generate spectra suitable for cycling photons on the X-B transition in CaF, and verify via Optical Bloch Equation simulations that one can achieve high photon scattering rates, which are important for fluorescent detection and laser cooling. Our method could offer significant simplifications in future experiments that would otherwise be prohibitively complex.
\end{abstract}
\maketitle

\section{Introduction}
Optical transitions are an indispensable part of the AMO experimental toolbox, as they are often used to manipulate and detect atoms and molecules. Substructure, such as hyperfine structure or spin-rotation structure, often leads to multiple optical transitions spaced by MHz to GHz. For many tasks, such as fluorescent detection and laser cooling~\cite{Hall1981,Chu1985}, it is necessary to simultaneously address these closely-spaced transitions in order to cycle photons. This becomes challenging when the number of desired frequencies is large, a situation that arises in complex molecules~\cite{Mitra2020,Baum2020} and atoms in large magnetic fields~\cite{Lu2010,Leefer2010,Mcclelland2006,Sukachev2010}.

To elaborate, even at zero magnetic field, optical cycling in molecules often requires multiple frequencies. Even in the case of a $^2\Sigma$ diatomic molecule with a single nuclear spin $I=1/2$, one would need four frequencies when the spin-rotation and hyperfine splittings are resolved in the ground electronic manifold~\cite{Shuman2010,Shelyazkova2014,Hemmerling2016}. In complex molecules with multiple nuclear spins or molecules with higher nuclear spin, the number of frequencies needed can become much larger. In atoms, a similar situation occurs at finite magnetic fields. Even in the case of an alkali atom, the ground and excited electronic manifolds split into a total of $4(2I+1)$ levels, where $I$ is the nuclear spin. One would generically need $2(2I+1)$ distinct frequencies in order to optically cycle photons.

Starting from monochromatic input light, one often uses an acousto-optic modulator (AOM) to shift the light to reach a desired target frequency. In this approach, the number of AOMs and associated optical components scales linearly with the number of required frequencies. Hence, this approach becomes impractical when there is a large number of frequencies. Additionally, AOMs have limited bandwidths and changes in their modulation frequencies are accompanied by changes in beam pointing, which is undesirable in many cases. Another common method to generate multiple frequencies is sinusoidal phase modulation, which can be accomplished by passing light through an electro-optic modulator (EOM). When the phase of monochromatic light is sinusoidally modulated, sidebands spaced by the modulation frequency are created about the carrier. Although there exist fortuitous cases where the resulting spectrum produces a near match to the desired set of target frequencies, invariably, undesired spectral components remain. 

By considering arbitary phase modulation profiles beyond sinusoidal ones, one can create arbitrary optical spectra. Nevertheless, it is not trivial to find the appropriate phase modulation profiles that will produce a target spectrum. In this work, we show that this inversion problem is equivalent to a computer-generated holography (CGH) problem, where well-known algorithms such as the Gerchberg-Saxton algorithm (GSA) exist~\cite{Gerchberg1972,Milster2020}. We show that our CGH-based method can be straightforwardly adapted for use in laser systems with non-linear optical processes such as sum frequency generation (SFG) and second harmonic generation (SHG). As a proof-of-principle demonstration, we generate an optical spectrum that can be used to cycle photons on the X-B transition in CaF molecules and show via Optical Bloch Equation (OBE) simulations that favorable photon cycling could be achieved with our method.

Outline: In Section \ref{SpecRHol}, we show how engineering a spectrum via phase modulation is equivalent to generating a 1D hologram using a phase-only spatial light modulator (SLM). In Section~\ref{GSAIntro}, we introduce GSA as a method to solve the spectral engineering problem, and show that its solutions are compatible with SFG and SHG. In Section \ref{linsys}, we discuss technical considerations for implementing our method. In Section \ref{PoP}, we present procedures to calibrate our system. We also show an example of GSA producing an optical spectrum that addresses the four $N=1$ hyperfine levels of CaF on the $\text{X}(N=1) \rightarrow \text{B}(N=0)$ optical cycling transition. In section \ref{addConc}, we discuss additional considerations when addressing atoms and molecules. In our method, the phase is modulated with a repeated waveform of arbitrary period. Since the finite excited state lifetimes in atoms or molecules provide additional timescales, waveforms with different periods could lead to different responses despite producing similar spectra. In Section \ref{conc}, we summarize our results and identify situations where our method could be particularly advantageous.

\section{Equivalence of Hologram Generation and Spectral Engineering via Phase Modulation}\label{SpecRHol}
In this section, we show that spectral engineering via phase modulation is equivalent to a well-known problem in optics -- how to generate a light intensity pattern $I(\tilde{x},\tilde{y})$ in the image plane ($\tilde{x}$-$\tilde{y}$ plane) by modulating only the light's phase in the Fourier plane ($x$-$y$ plane). For concreteness, consider an imaging system consisting of a single lens with focal length $f$. Suppose an electric field of the form $E(x,y) =E_0 e^{i \phi(x,y)}$ is in the Fourier plane. In the image plane (under the paraxial approximation), the electric field is given by

\begin{equation}
E(\tilde{x},\tilde{y} ) = \frac{e^{i 4\pi f}}{i} \int dx dy \, E_0 e^{i\phi(x,y)} e^{-i k_0 (x \tilde{x} + y \tilde{y})/f},
\end{equation}
where $k_0 = 2\pi/\lambda$, and $\lambda$ is the wavelength of the light. Thus, the electric field $E(\tilde{x},\tilde{y})$ in the image plane is the Fourier transform of the electric field $E(x,y)$ in the Fourier plane, a well-known result in Fourier optics.

In CGH, given a desired target intensity pattern $I(\tilde{x},\tilde{y})$ in the image plane, one seeks the optimal 2D phase profile $\phi(x,y)$. Explicitly, one tries to invert
\begin{equation}
I(\tilde{x},\tilde{y} ) =\left| \int dx dy \, E_0 e^{i\phi(x,y)} e^{-i (x \tilde{x} + y \tilde{y})} \right|^2,
\end{equation}
where we have picked units such that $k_0$ and $f$ are 1. In 1D, the corresponding equation to invert is
\begin{equation} \label{1DHolProb}
I(\tilde{x})= \left|\int dx\; E_0 e^{i\phi(x)}e^{-i \tilde{x} x}\right|^2,
\end{equation}
where one seeks an optimal $\phi(x)$ when given $I(\tilde{x})$.

The 1D hologram problem is in fact equivalent to the problem encountered in engineering spectra. When creating spectra via phase modulation, one attempts to produce a target power spectral density $S(\omega)$ by applying a temporally varying phase $\phi(t)$ to a monochromatic light field $E_0 e^{i \omega_0 t}$. The problem is thus to invert
\begin{equation}
\label{SpecEngProb1}
S(\omega)=\left| \int dt\; E_0 e^{i(\omega_0 t+\phi(t))}e^{-i\omega t}\right|^2.
\end{equation}
Identifying time (frequency) with the spatial coordinate in the Fourier (image) plane ($ \{x,\tilde{x}\} \rightarrow \{t,\omega\} $), we see that Eq.~(\ref{1DHolProb}) and Eq.~(\ref{SpecEngProb1}) are identical up to a frequency offset of $\omega_0$ in $\omega$. Therefore, finding a temporal phase modulation profile $\phi(t)$ that produces a desired spectrum $S(\omega)$ is equivalent to finding a spatial phase profile $\phi(x)$ that produces a target image $I(\tilde{x})$ in 1D hologram generation. Hence, algorithms developed for CGH can directly be used to engineer spectra via phase modulation.

\section{Spectral Engineering with Computer Holography Solutions} \label{GSAIntro}
In order to find a phase modulation profile $\phi(t)$ that produces a target spectrum $S(\omega)$, one has to invert Eq.~(\ref{SpecEngProb1}). While many algorithms have been developed for this purpose~\cite{Fienup1982, Fienup1986, Wang2017}, in this work, we will use GSA \cite{Gerchberg1972}, a well-known iterative Fourier transform algorithm (IFTA) that produces approximate solutions. 

\begin{figure}[th]
{\includegraphics[width=\columnwidth]{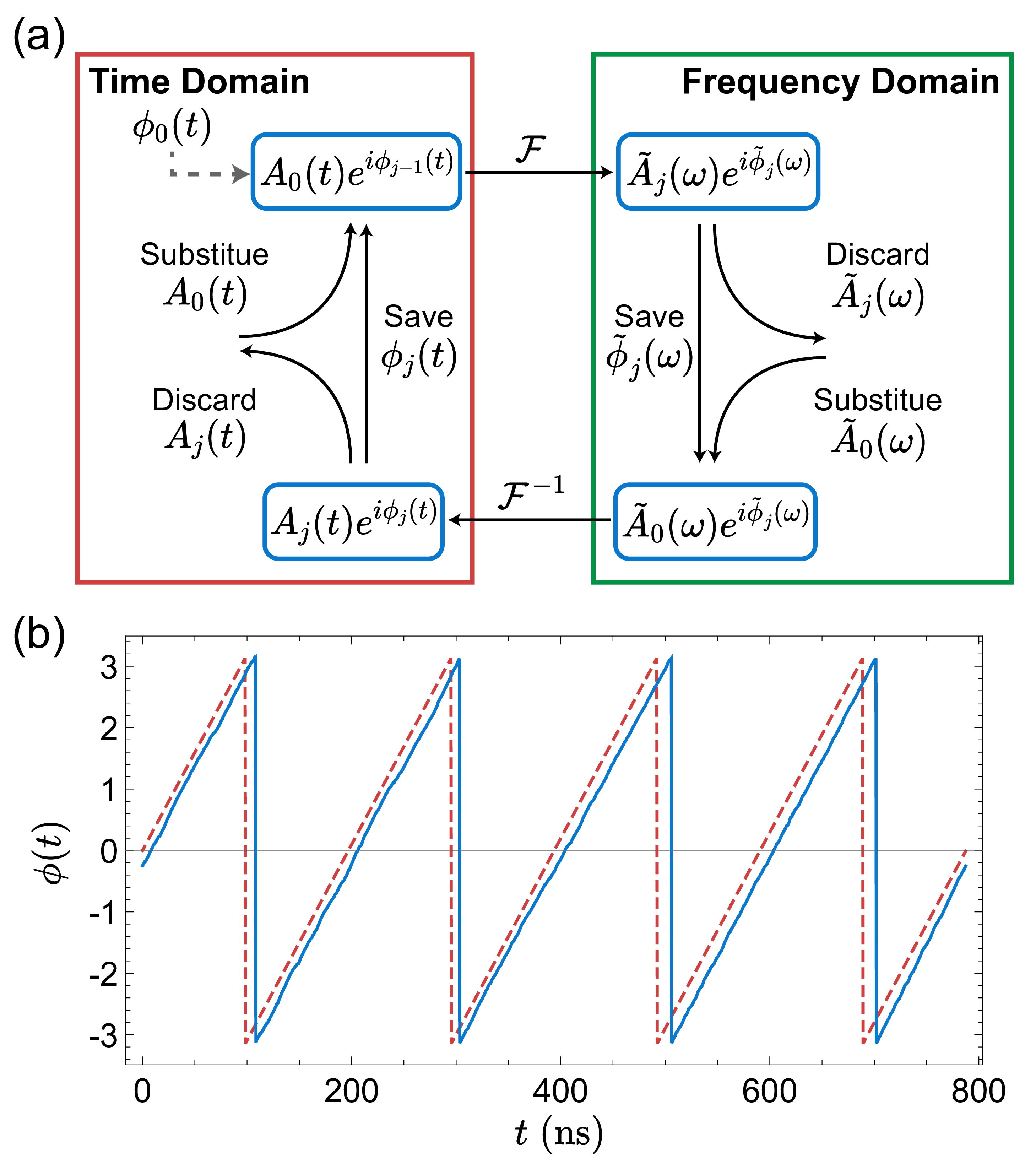}}
\caption{\label{fig:GSALoop} The Gerchberg-Saxton Algorithm (GSA). (a) GSA is an iterative algorithm that enforces amplitude constraints on a signal and its Fourier transform. In spectral engineering, the amplitudes in the time and frequency domains are constrained via $A_0(t)$ and $\tilde{A}_0(\omega)$, respectively. (b) A GSA solution (blue solid) for a Lorentzian target (+5\,MHz from carrier, 100 kHz FWHM) closely resembles a serrodyne solution (+5\,MHz, red dashed).}
\end{figure}

We describe GSA in brief (Fig.~\ref{fig:GSALoop}a). Given a target amplitude $\tilde{A}_0(\omega)$ in the frequency domain and an input amplitude $A_0(t)$ in the time domain, the algorithm returns a phase modulation profile $\phi(t)$ that closely reproduces $\tilde{A}_0(\omega)$. Starting with a seed phase $\phi_0(t)$, one follows the following steps:
\begin{enumerate}
\item The time domain signal $A_0(t)e^{i\phi_{j}(t)}$ is propagated to the frequency domain via a Fourier transform, producing $\mathcal{F}[A_0(t)e^{i\phi_{j}(t)}]=\tilde{A}_{j+1}(\omega)e^{i\tilde{\phi}_{j+1}(\omega)}$, expressed in terms of a propagated amplitude $\tilde{A}_{j+1}(\omega)$ and phase $\tilde{\phi}_{j+1}(\omega)$.
\item The propagated amplitude is discarded and replaced with the target amplitude $\tilde{A}_0(\omega)$ to generate the frequency domain signal $\tilde{A}_0(\omega)e^{i\tilde{\phi}_{j+1}(\omega)}$. This step enforces a constraint on the amplitude in the frequency domain.
\item The frequency domain signal is propagated back to the time domain via an inverse Fourier transform, producing $\mathcal{F}^{-1}[\tilde{A}_0(\omega)e^{i\tilde{\phi}_{j+1}(\omega)}]=A_{j+1}(t)e^{i\phi_{j+1}(t)}$.
\item The propagated amplitude $A_{j+1}(t)$ is discarded and replaced with the input amplitude $A_0(t)$ acting as a constraint, producing the subsequent time domain signal $A_0(t)e^{i\phi_{j+1}(t)}$.
\end{enumerate}
The algorithm terminates either after a predetermined number of steps or when $|\tilde{A}_j(\omega)-\tilde{A}_0(\omega)|<\varepsilon$ for a specified convergence tolerance $\varepsilon$.

\subsection{Creating a Single Frequency}
Although CGH algorithms such as IFTAs often appear as ``black boxes" whose solutions elude physical intuition, GSA solutions for simple target spectra can often be interpreted. A simple case concerns shifting the angular frequency of an input light field $E_0 e^{i\omega t}$ by $\delta$, $E_0 e^{i\omega_0t} \rightarrow E_0 e^{i(\omega_0+\delta)t}$. It is clear that $\phi(t) = \delta t+\Phi_0$ is a solution, where $\Phi_0$ is a constant that does not affect the power spectrum. As shown in Fig.~\ref{fig:GSALoop}b, GSA reproduces this solution, modulo $2\pi$. This sawtooth-like solution, the so-called ``serrodyne" solution, has been used to shift frequencies both in the microwave \cite{Cumming1957} and optical domains~\cite{Wong1982,Johnson2010}. Note that since the constant $\Phi_0$ does not affect the power spectral density, it is not unique and depends on the initial seed $\phi_0(t)$. Hence, as in hologram generation, there always exist multiple $\phi(t)$ yielding identical spectra.

\subsection{Creating Multiple Frequencies}
Simple solutions to generate multiple frequencies are not always available. One approach involves stitching together multiple serrodyne waveforms to create multiple frequencies~\cite{Rogers2011}. The desired power of each frequency component can be controlled via the duration of each serrodyne. However, concatenation introduces imperfections, since the transitions between different serrodynes contribute additional frequency content. This content becomes particularly problematic when the duration of each serrodyne is short. We will show that GSA solves this problem by smoothly stitching together serrodynes.

We note in passing that some simple solutions become obscured when using GSA. It is well-known that modulation with a sinusoidal drive $\phi(t)=\beta \cos(\Omega t)$ generates multiple sidebands spaced by $n\Omega$ from the carrier, with strengths $|J_n(\beta)|^2$, where $\beta$ is the modulation depth and $J_n(x)$ is the $n^{\text{th}}$ order Bessel function of the first kind:
\begin{equation}\label{ODEOM}
\Bigg|\int dt\;e^{i\beta\cos(\Omega t)}e^{-i\omega t}\Bigg|^2 \propto \sum_{n=-\infty}^{\infty} |J_n(\beta)|^2\delta(\omega-n\Omega).
\end{equation}
\noindent
Suppose the target is exactly the spectrum created by sinusoidal modulation with a depth $\beta >\pi$. Due to the restriction of a GSA solution to $\phi(t)\in[-\pi, \pi)$, the simple sinusoidal solution would not be apparent.

\subsection{Use in Non-Linear Optical Processes}
The CGH approach can be applied even in the presence of non-linear optical processes such as SFG and multiple harmonic generation. In SFG, two monochromatic light fields $E_1 e^{i\omega_1 t} $ and $E_2 e^{i\omega_2 t}$ combine to produce a new field proportional to $ E_1 E_2 e^{i(\omega_1+\omega_2)t}$. Since the second light field only acts to modulate the amplitude and frequency-shift the first light field by $\omega_2$, a spectrum centered about $\omega_1+\omega_2$ can be generated by creating the same spectrum on the first light field but centered about $\omega_1$. 

For $n^{\text{th}}$ harmonic generation, an input of $E_0 e^{i\omega_0 t}$ produces an output proportional to $E_0^n e^{i n\omega_0 t}$. Phase-modulating the input light by $\phi(t)$ produces a phase modulation of $n\phi(t)$ on the output light. Thus, a desired spectrum $S(\omega)$ in the $n^{\text{th}}$ harmonic can be generated by applying $\phi(t)=\phi_{\text{soln}}(t)/n$ to the input light, where $\phi_{\text{soln}}(t)$ is the CGH solution.

The compatibility with non-linear processes provides practical advantages. Often, high power visible light can be created via SFG or SHG applied to near-infrared light, where high power fiber amplifiers and high-bandwidth fiberized EOMs are available. In Section~\ref{PoP}, we provide a proof-of-principle demonstration of such an implementation, where a target spectrum centered about 531\,nm is produced by phase-modulating light at 1062\,nm.

\section{Technical Considerations} \label{linsys}
In this section, we discuss technical considerations when implementing arbitrary phase modulation for spectral engineering. We generate a phase modulation signal $\phi(t)$ by applying a voltage $V(t)$ to an EOM. The advent of high-bandwidth fiberized EOMs, available with bandwidths up to 10's of GHz, makes this a feasible approach for generating optical spectra spanning 100s of MHz. Nevertheless, various effects make the system non-ideal, leading to behavior deviating from Eq.~(\ref{SpecEngProb1}). These effects can arise at the level of $V(t)$ (electronic) or after the phase $\phi(t)$ has been imprinted by the EOM on the light field (optical). The electronic effects can be separated into linear ones and nonlinear ones. Effects that are linear in the voltage signal $V(t)$ can be described by a transfer function. Once the transfer function is known, it can be compensated. Electronic effects that are non-linear are much more difficult to deal with. Our strategy is to minimize these effects. Beyond electronic effects, a major possible source of non-ideal behavior is chromatic dispersion, which can be large especially when long optical fibers are used. We will show that as long as the dispersion is linear, our method is not affected.

\subsection{Linear Electronic Effects}
Certain electronic effects encountered when generating $V(t)$, such as finite bandwidth and frequency-dependent amplitude responses in various components, modify the resulting phase modulation waveform from $\phi(t)$ to $\phi_{\text{eff}}(t)$. As long as these effects are linear, they can be captured by a transfer function acting in the frequency domain. Specifically, the resulting phase modulation in the frequency domain, $\tilde{\phi}_{\text{eff}}(\omega) = \int dt \,\phi_{\text{eff}} (t) e^{-i\omega t}$, is given by $\tilde{\phi}_{\text{eff}}(\omega)= T(\omega) \tilde{\phi}(\omega)$, where $T(\omega)$ is the transfer function and $\tilde{\phi}(\omega)$ is the Fourier transform of the initial signal. The transfer function $T(\omega)$ can be expressed as $T(\omega) = A(\omega) e^{i \Phi(\omega)}$, where $A(\omega)$ is the amplitude response and $\Phi(\omega)$ is the phase response. We will show in section~\ref{PoP} how to measure both of these functions. We note in passing that the hologram analogs to these linear effects include non-ideal SLM surface flatness and spatially dependent transmission of the SLM when it is placed in the Fourier plane.

\subsection{Non-Linear Electronic Effects} 
Non-linear electronic effects can occur when generating a time-dependent voltage $V(t)$. For example, suppose $V(t)$ consists of a single tone at $\Omega$. Amplifier non-linearities will produce higher harmonics at $n \Omega$. This situation worsens when $V(t)$ contains multiple frequencies, since intermodulation distortion produces signals at the sums and differences of the input frequencies. Since non-linear effects cannot be captured by a transfer function, we try to minimize these effects through an appropriate choice of components. Specifically, in our implementation, we impose a high frequency cutoff of 500\,MHz in $\phi(t)$, and verify using a spectrum analyzer that for single frequency inputs, harmonics in $V(t)$ are suppressed by at least 40\,dB.

\subsection{Chromatic Dispersion}
Beyond electronic effects that modify the phase $\phi(t)$, optical effects could also lead to non-ideal behavior. In laser systems that contain optical fibers, chromatic dispersion is an effect that one has to consider. Here, we show that as long as the material has linear chromatic dispersion over the modulation bandwidth, dispersion only introduces a global time delay and our method is unaffected. 

Suppose a material has linear chromatic dispersion, i.e. its index of refraction is approximately described by $n(\lambda) \approx n_0 + \alpha (\lambda-\lambda_0)$, where $\lambda$ is the wavelength, $n_0$ is the index of refraction at $\lambda_0$, and $\alpha = \frac{dn}{d\lambda}$. After propagation of length $L$, an input light field with Fourier spectrum $\tilde{E}(\omega)=\int dt \,E(t) e^{-i\omega t}$ becomes
\begin{align}
E_{\text{out}}(t) &= \int \frac{d\omega}{2\pi}\, \tilde{E}(\omega)e^{-in(\lambda)\frac{2\pi}{\lambda}L}e^{i\omega t}\nonumber\\
&=e^{-i2\pi\alpha L}\int \frac{d\omega}{2\pi}\, \tilde{E}(\omega)e^{-i(n_0-\alpha \lambda_0)\frac{\omega}{c}L}e^{i\omega t}
\end{align}
Now, suppose the input light is monochromatic with frequency $\omega_0$ and is phase-modulated by $\phi(t)$, i.e. $E(t) = E_0 e^{i(\omega_0 t +\phi(t))}$. After propagation, the output light field is given by
\begin{align}
E_{\text{out}}(t)&=E_0 e^{-i2\pi\alpha L}\int \frac{d\omega}{2\pi}\int dt' \, e^{i(\omega_0 t'+\phi(t'))}\nonumber \\
& \qquad\qquad\qquad e^{-i( t'-t+(n_0-\alpha\lambda_0)L/c)\omega}\nonumber\\
& =E_0 e^{-i \omega_0 n_0 L/c} e^{i(\omega_0t +\phi(t-\Delta t))},
\label{linDisp}
\end{align}
where $\Delta t = (n_0-\alpha\lambda_0)L/c$. Therefore, linear dispersion leads to a trivial phase factor $e^{-i \omega_0 n_0 L/c}$ and a time delay $\Delta t$ in the phase modulation $\phi(t)$, which does not affect the spectrum. In the case of optical fibers, linear chromatic dispersion is a good approximation over $\sim\text{GHz}$ bandwidths. 

\section{Proof-of-Principle Demonstration} \label{PoP}
In this section, we present methods to measure the amplitude and phase response, $A(\omega)$ and $\Phi(\omega)$, of the transfer function. Compensation can then be applied by dividing the intended drive by the transfer function in the frequency domain. We will show that when compensation is applied, we obtain good agreement between observed and simulated spectra.

\begin{figure}[t]
{\includegraphics[width=\columnwidth]{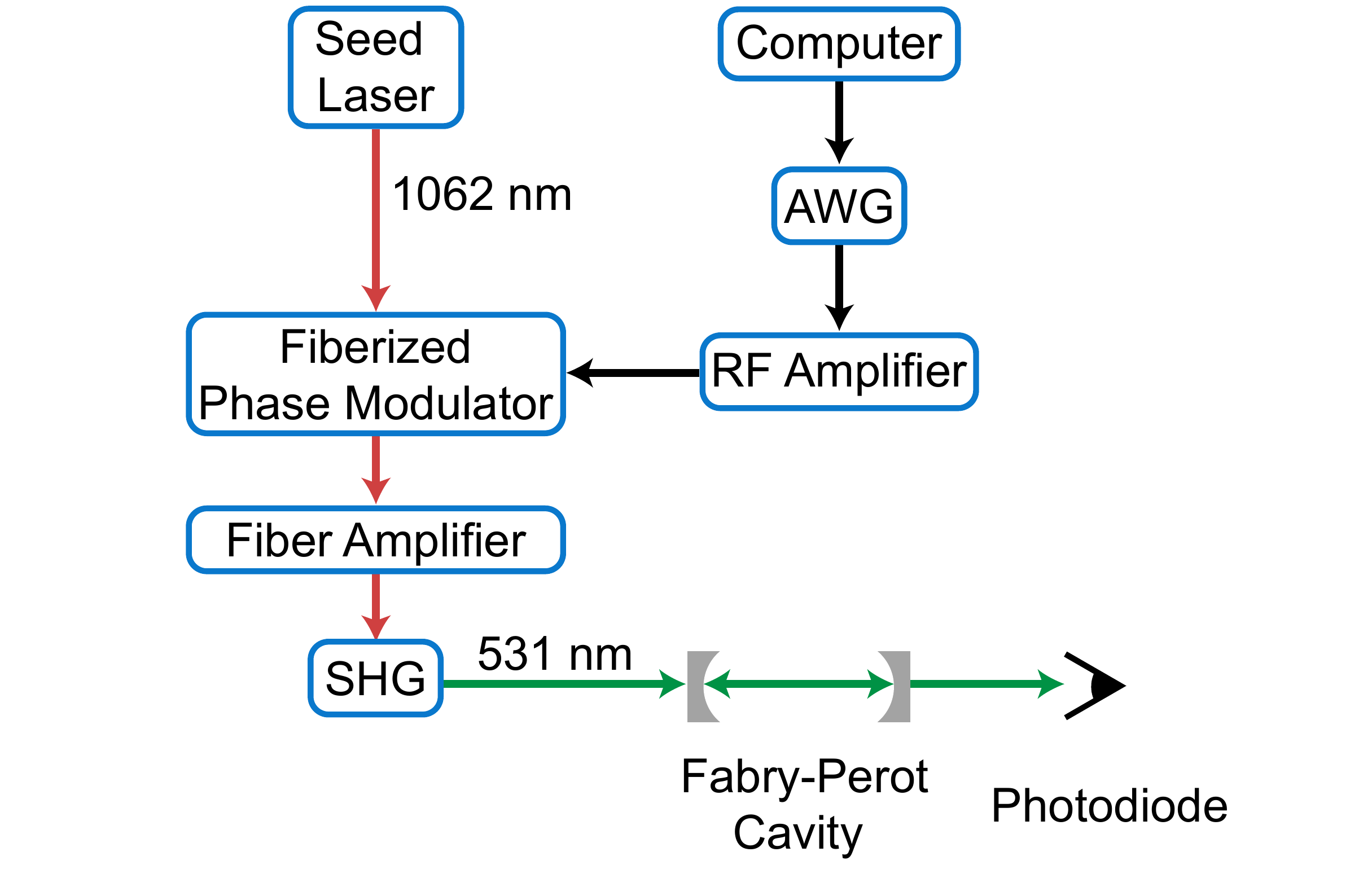}}
\caption{\label{fig:SysLayout} Experimental Setup. 1062\,nm light from a seed laser is sent through a fiberized phase modulator, amplified, and then frequency doubled (SHG) to 531\,nm. The phase applied by the modulator is controlled by the amplified output of an AWG. To measure the spectrum of the 531\,nm light, we observe the tranmission through a scanning Fabry-Perot cavity.}
\end{figure}

Our system (Fig. \ref{fig:SysLayout}) consists of a seed laser at 1062\,nm whose light passes through a fiberized phase modulator (2\,GHz bandwidth) that is controlled by the amplified output of an arbitrary waveform generator (AWG). 
After phase modulation, the light enters a high-power fiber amplifier, whose output is frequency doubled to 531\,nm. To characterize the spectrum of the doubled light, we measure the transmission $\mathcal{T}$ through a Fabry-Perot cavity (1.5\,GHz free spectral range, 5.8(1)\,MHz full-width-half-max linewidth) as a function of the cavity length, which is scanned using a piezo-electric actuator. By converting the piezo drive voltage to an angular frequency difference $\delta$ and aligning $\delta=0$ with the carrier, one produces a transmission curve $\mathcal{T}(\delta)$, which is directly proportional to the power spectrum of the light $S(\omega)$, with $\delta =\omega$. To calibrate the piezo response, we apply sinusoidal phase modulation to the light at frequency $\Omega$, which produces sidebands spaced by $n\Omega$ about the carrier. We approximate the piezo response to be linear in the drive voltage. Piezo non-linearities lead to small deviations, which are unimportant for our present work. 

\subsection{Calibrating the Amplitude Response}
\begin{figure}[t]
{\includegraphics[width=\columnwidth]{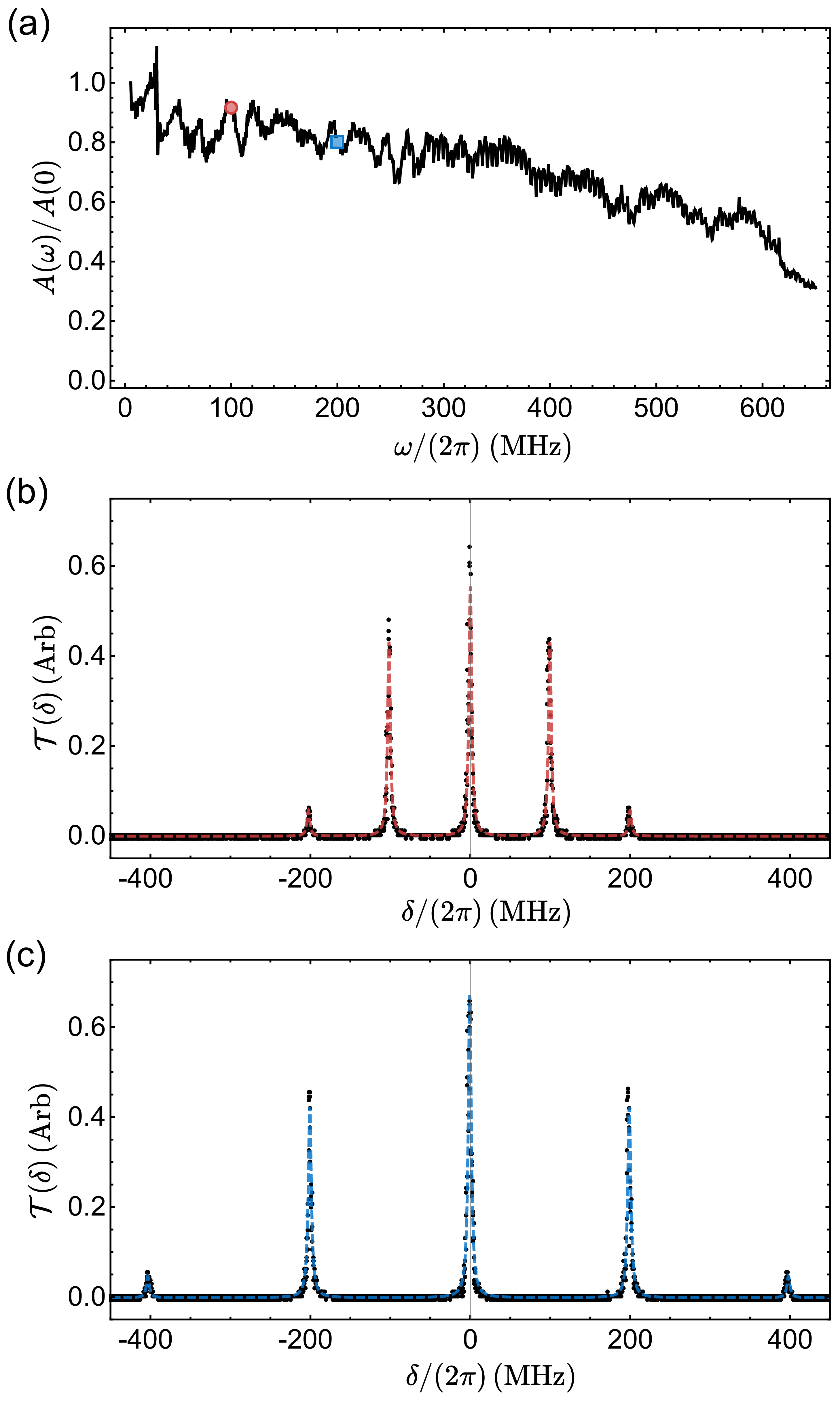}}
\caption{\label{fig:amp} Amplitude Response. (a) Measured amplitude transfer function $A(\omega)/A(0)$ obtained by fitting cavity transmission spectra of sinusoidally modulated light. Example transmission spectra $\mathcal{T}(\delta)$ (black points) and their fits are shown in (b) 100\,MHz (red dashed) and (c) 200\,MHz (blue dashed). These frequencies are marked in (a) by the red circle and blue square, respectively.
}
\end{figure}

To measure the amplitude response $A(\omega)$, we apply a voltage to our EOM of the form $V(t)=V_0 \cos(\Omega t)$ with frequencies $\Omega/(2\pi)$ ranging from 5 to 650\,MHz. This creates a phase modulation profile of $\phi(t) = \beta \cos(\Omega t)$, which produces sidebands at $n\Omega$ on the undoubled light. The effect of frequency-doubling can be absorbed into the modulation depth via $\beta \rightarrow 2\beta$. We measure the resulting optical spectra of the doubled light using a scanning Fabry-Perot cavity and fit the results to the known sinusoidal modulation solution (Eq.~(\ref{ODEOM})), extracting the effective modulation depth $\beta_{\text{eff}}$. This provides the amplitude transfer function $A(\omega)$ (Fig.~\ref{fig:amp}). The calibration is performed such that the measured modulation depth stays near $\beta_{\text{eff}}\approx1.2$, producing only a few sidebands whose amplitudes are sensitive to variations in $\beta_{\text{eff}}$. 

\subsection{Calibrating the Phase Response}
To measure the phase response $\Phi(\omega)$, we developed an interferometric method that is sensitive to the relative phase shift between two frequencies. Inspired by the fact that a non-symmetric spectrum is produced when the modulation signal $\phi(t)$ breaks time-reversal symmetry, we apply modulation of the form
\begin{equation} \label{f2fPhase}
\phi(t)=\beta_1\cos(\Omega t)+\beta_2\cos(2\Omega t+\varphi).
\end{equation}
One observes that unless $\varphi =0$ modulo $\pi$, $\phi(t)$ is not time-reversal symmetric.
Incorporating the amplitude and phase response of the system, the effective modulation signal is then
\begin{align} \label{f2fPhaseEff}
\phi_{\text{eff}}(t)=& A(\Omega)\beta_1\cos(\Omega t+\Phi(\Omega))\nonumber\\
&+A(2\Omega)\beta_2\cos(2\Omega t+\Phi({2\Omega})+\varphi).
\end{align}
Shifting the time coordinate by $t\rightarrow t-\Phi(\Omega)/\Omega$, one finds that
\begin{align} \label{f2fPhaseEffPhase}
\phi(t)=& A(\Omega)\beta_1\cos(\Omega t)\nonumber\\
&+A(2\Omega)\beta_2\cos(2\Omega t+\Delta \Phi(\Omega)+\varphi),
\end{align}
where $\Delta \Phi(\Omega) = \Phi (2\Omega) - 2\Phi (\Omega)$. The sidebands generated by the $2\Omega$ modulation will interfere with the sidebands generated by the $\Omega$ modulation, and the resulting spectrum encodes information about the relative phase $\Delta \Phi (\omega)$. Applying Eq.~(\ref{ODEOM}), one finds that the cavity spectrum $S(\omega)$ is given by
\begin{eqnarray}\label{f2fSpec}
& &\Bigg|\sum_{m}J_{n-2m}(\beta_{1,\text{eff}})J_m(\beta_{2,\text{eff}})e^{in\pi/2}e^{-im(\Delta \Phi+\varphi+\pi/2)}\Bigg|^2
\nonumber \\
& & \quad \times \, \, \, \delta(\omega_0-n\Omega),
\end{eqnarray} 
where $\beta_{1,\text{eff}}= A(\Omega)\beta_1$ and $\beta_{2,\text{eff}}= A(2\Omega)\beta_2$.

\begin{figure}[h!]
\includegraphics[width=\columnwidth]{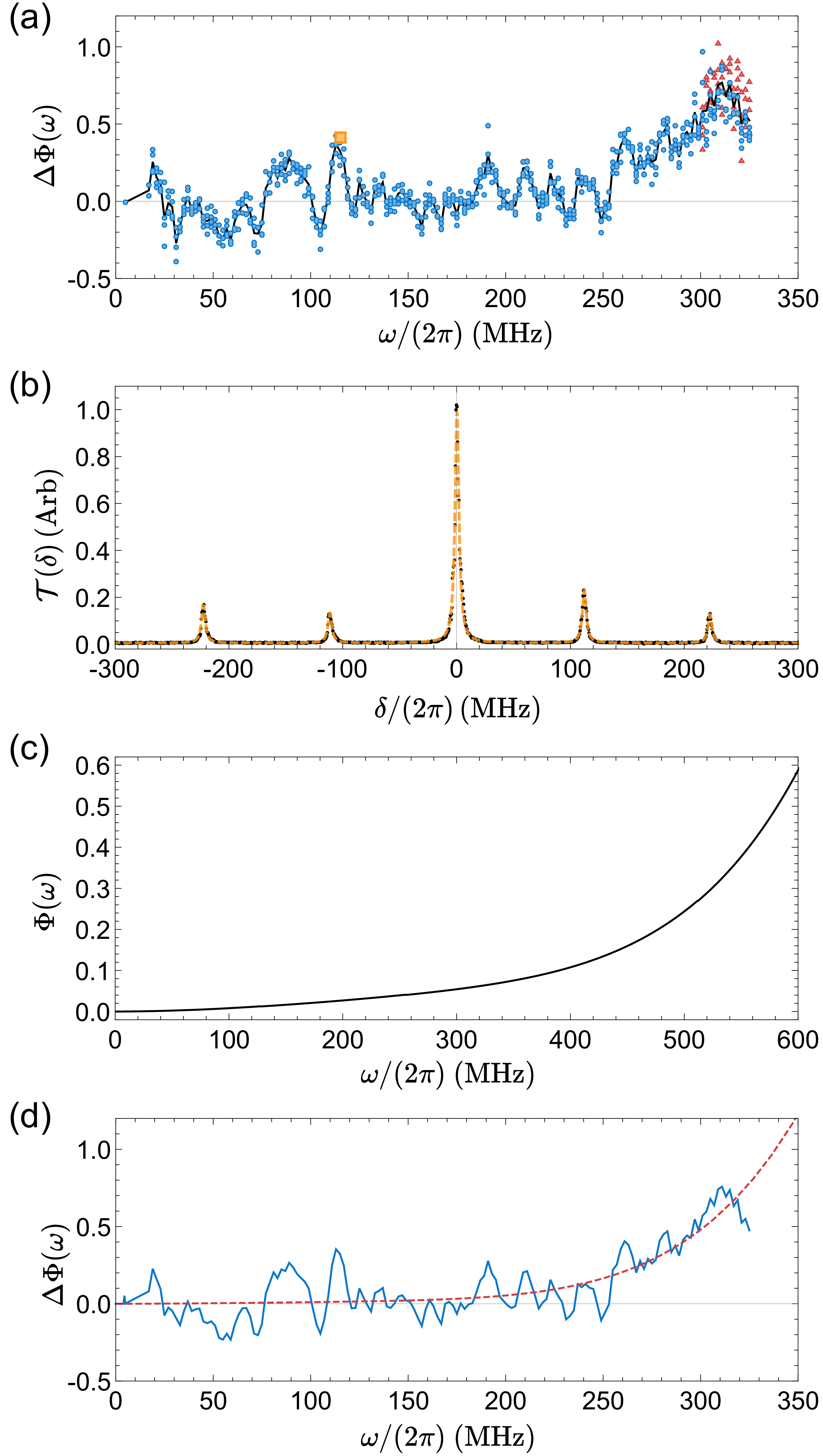}
\caption{ Phase Response. (a) The measured relative phase response $\Delta \Phi(\omega) = \Phi (2\omega) - 2\Phi (\omega)$ extracted from modulation with $\varphi=0$ (blue circles) and $\varphi=\pi/2$ (red triangles). Shown in black is the rolling average. The two sets of $\varphi$ data are plotted in the frequency ranges where they are sensitive to $\Delta\Phi(\omega)$. (b) Cavity scan (black dots) and the fitted curve (orange dashed) of a representative interferometric spectrum at the frequency indicated by the orange square in (a). (c) Phase response $\Phi(\omega)$ (black solid) reconstructed from the measured $\Delta \Phi (\omega)$. (d) Comparison between the measured relative phase $\Delta \Phi(\omega)$ (blue solid) and the relative phase obtained from the reconstructed phase response $\Phi(\omega)$ (red dashed).}
\label{fig:phase}
\end{figure}

Our procedure to obtain $ \Phi (\omega)$ is as follows. First, we calibrate our amplitude response as described above. We then set the modulation depths to $\beta_{1,\text{eff}}=\beta_{2,\text{eff}}=0.7$ and measure the resulting spectra for $\varphi=0$ and $\varphi=\pi/2$. The amplitudes are chosen to maximize contrast while keeping the spectra relatively simple. The two values of $\varphi$ help resolve a quadrant ambiguity in Eq.~(\ref{f2fSpec}). Fig.~\ref{fig:phase}a shows the extracted relative phase $\Delta \Phi (\omega)$ and Fig.~\ref{fig:phase}b shows a representative fitted spectrum.

To reconstruct $\Phi(\omega)$ from $\Delta\Phi(\omega)$, we first make the following assumptions. We assume that $\Delta \Phi(\omega) \approx 0$ at the lowest measured frequency. We also assume that the leading order term in $\Phi(\omega)$ is at most quadratic. We rule out a linear leading order term because of the following. A phase delay that is linear in frequency translates to a common time delay, which is not observable in our interferometric measurement. We further assume that $\Phi(\omega)$ is smooth and slowly varying over the frequency range that we measure. We therefore fix $\Phi(0) = 0$ and fit the $\Delta \Phi(\omega)$ data to a low-order polynomial with zero offset and vanishing linear term. The fitted and smoothed version of $\Delta \Phi(\omega)$ is then used to reconstruct $\Phi(\omega)$.

To obtain the total phase response $\Phi(\omega)$, we start near $\omega=0$ and proceed in a ladder-like manner. For example, $\Phi(2\Omega)$ is obtained from lower frequency measurements using $\Phi(2\Omega)=\Delta\Phi(\Omega) + 2\Phi (\Omega)$. To verify the validity of our procedure, we obtain the relative phase $\Delta \Phi(\omega)$ from the reconstructed total phase $\Phi(\omega)$ (Fig.~\ref{fig:phase}c) and compare it with the measured relative phase $\Delta\Phi(\omega)$. We find good agreement, indicating proper reconstruction of $\Phi(\omega)$ from the measurements (Fig.~\ref{fig:phase}d).

\subsection{Compensated Signal Comparison}
\begin{figure}[t]
{\includegraphics[width=\columnwidth]{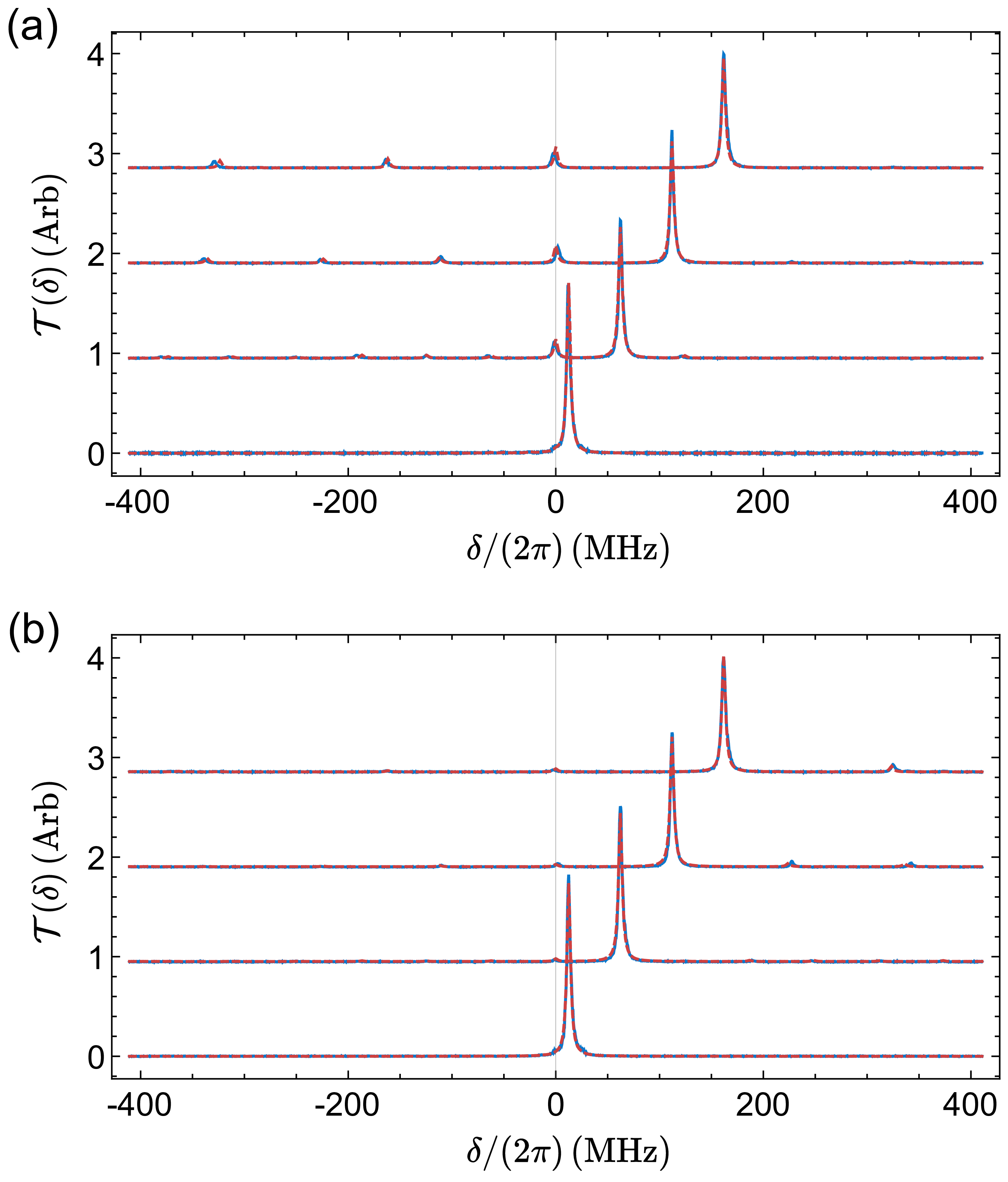}}
\caption{ Compensated and Uncompensated Serrodyne Spectra. (a) Measured (blue solid) and simulated spectra (red dashed) of uncompensated serrodynes, for serrodyne frequencies of 12.5\,MHz, 63.5\,MHz, 145\,MHz, and 165\,MHz. (b) Measured (blue solid) and simulated (red dashed) spectra of compensated serrodynes for the same serrodyne frequencies. For the simulated spectra, we apply a single vertical scaling factor, which is obtained by comparing the maxima of the simulated and measured spectrum for the uncompensated 12.5\,MHz serodyne. The cavity traces are shifted horizontally to align the main peaks with the simulation, and a single horizontal scaling factor is used for all traces. }
\label{fig:serroComp}
\end{figure}

To verify the accuracy of our measured transfer function, we compare measured and simulated spectra. The measured spectra come from compensated and uncompensated serrodyne waveforms, and are measured using our Fabry-Perot cavity. The simulated spectra come from applying the measured modulation transfer function to the phase modulation profiles and convolving with the measured Fabry-Perot lineshape (Fig.~\ref{fig:serroComp}).

To generate the compensated waveforms, we apply the inverse of the measured transfer function to the uncompensated modulation profiles. The frequency range of our serrodynes is between 12.5 and 190\,MHz, with the highest frequency chosen so that all serrodynes contain at least two compensated Fourier components. Apart from small deviations in the locations of the peaks that we attribute to a nonlinear piezo response, we find good agreement between measured and simulated spectra. This verifies that the measured transfer function is accurate and validates our transfer function measurement protocol.

We find that compensating for the transfer function leads to improved spectra. For example, the -1 order sideband and the carrier are suppressed on average to 24(8)\% and 16(10)\% of their respective values in uncompensated spectra. Furthermore, the 1st order sideband increases on average by 6(5)\%, suggesting improved efficiency. Nevertheless, even with compensation, the overall modulation efficiency, as measured by the power in the desired sideband, decreases with increasing serrodyne frequency. For (un)compensated modulation profiles, the efficiency decreases from (88)90\% at 12.5\,MHz to (48)50\% at 190\,MHz. While the improvements in efficiency are modest, compensation greatly suppresses spurious peaks within the modulation bandwidth, as shown in Fig.~\ref{fig:serroComp}.

\subsection{Proof-of-Principle Demonstration}
\begin{figure}[t]
{\includegraphics[width=0.95\columnwidth]{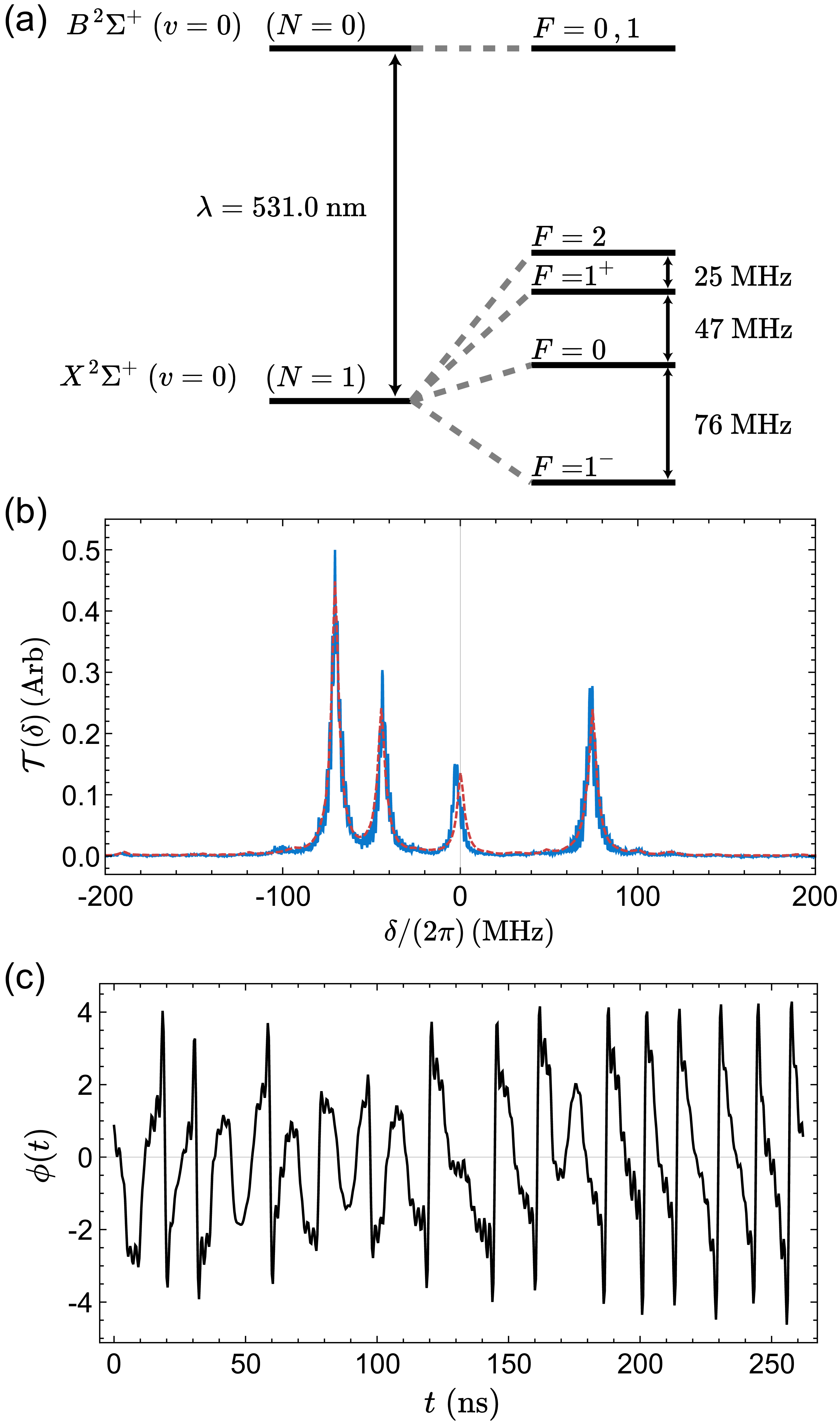}}
\caption{\label{fig:CaFSpec} Proof-of-Principle Spectrum Generation for $X(N=1)\rightarrow B(N=0)$ Transition in CaF. (a) Hyperfine structure of CaF for states in the $X(N=1)\rightarrow B(N=0)$ optical cycling transition. The excited hyperfine structure is unresolved. (b) A spectrum that can address all the ground hyperfine states in (a) is generated using GSA, and compensated by the measured transfer function. The hyperfine components are addressed with relative powers $5:3:1:3$. Shown in blue solid (red dashed) is the measured (simulated) cavity transmission $\mathcal{T}(\delta)$. The horizontal scaling is identical to that in Fig.~\ref{fig:serroComp}. A horizontal shift is applied to overlap the measured and simulated maxima ($F=2$ feature). (c) Compensated $\phi(t)$ used to generate the spectrum in (b).}
\end{figure}

As a proof-of-principle demonstration, we use the above methods to generate a spectrum relevant to laser-cooling CaF \cite{Anderegg2017,Truppe2017,Anderegg2018}. In CaF, one can cycle photons by addressing the $X(N=1) \rightarrow B(N=0)$ optical transition at 531\,nm. The ground state $X(N=1)$ is composed of four hyperfine states while the excited state hyperfine structure is unresolved (Fig.~\ref{fig:CaFSpec}a). As a result, four optical frequencies are needed to prevent molecules from falling into unaddressed ``dark'' hyperfine manifolds. We generate the required spectrum with the appropriate splittings using GSA. The intensities are picked with weights corresponding to the degeneracies of the hyperfine manifolds. As shown in Fig.~\ref{fig:CaFSpec}b, the measured spectrum generated using GSA and the simulated target spectrum agree well, demonstrating that our method can be used to generate multi-frequency spectra for typical hyperfine spacings. The good agreement also shows that our method works in SHG systems and that chromatic dispersion in long fibers is not a concern.

\subsection{Comparison of GSA Solutions with Concantenated Serrodynes}
\begin{figure}[t]
{\includegraphics[width=\columnwidth]{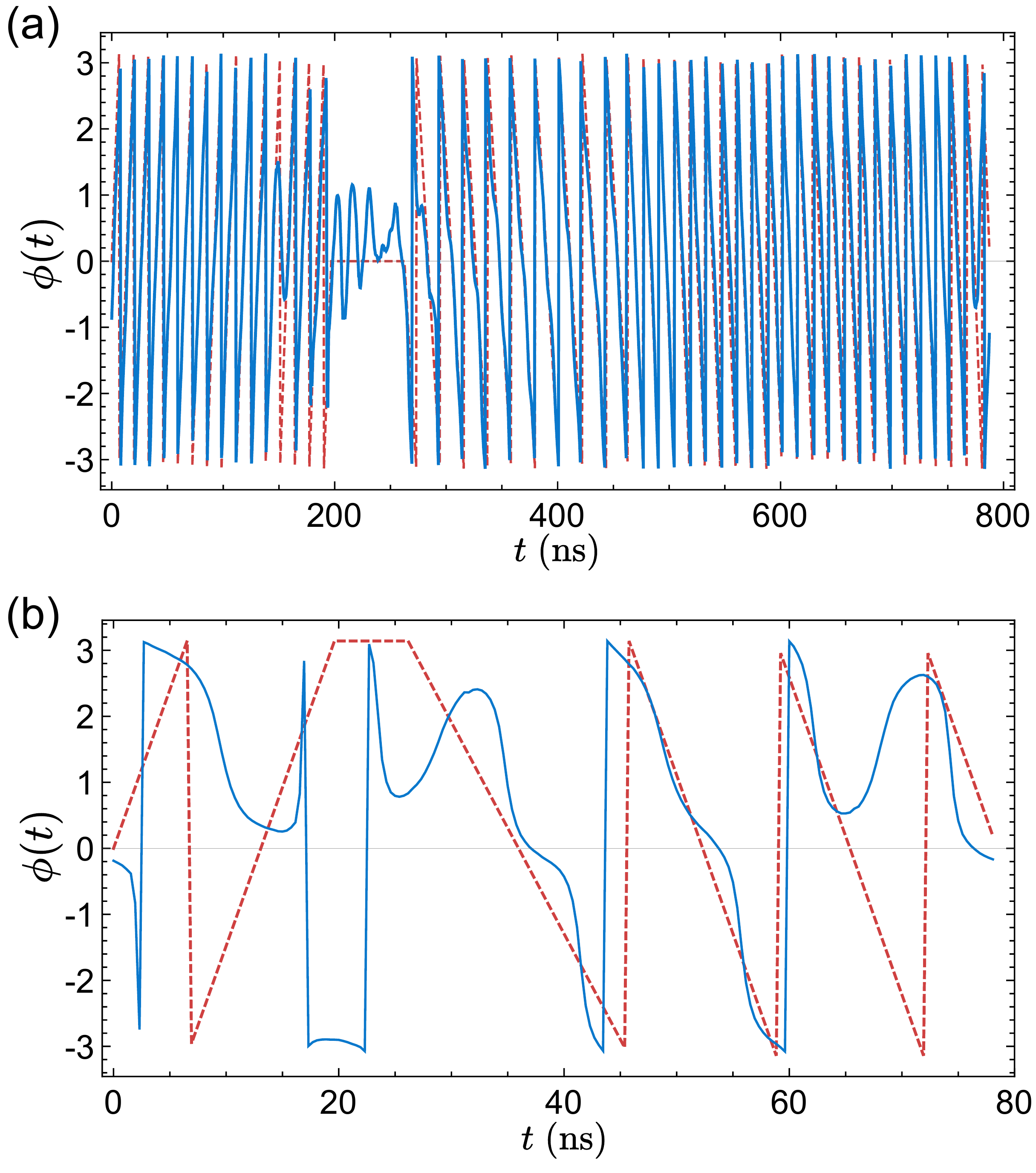}}
\caption{Comparison of Concatenated Serrodynes and GSA Solutions. The target spectrum is the same as in Fig.~\ref{fig:CaFSpec}. (a) Shown are the seeded GSA solution $\phi(t)$ (bue solid) and the initial seed $\phi_0(t)$ (red dashed) for a waveform duration of $\tau=787\,\text{ns}$. This waveform duration yields a Fourier limit $1/(2\pi\tau) = 200\,\text{kHz}$, and the smallest hyperfine spacing of 25\,MHz is therefore fully resolvable. (b) For a shorter waveform duration of 78.7\,\text{ns}, the corresponding Fourier limit is $1/(2\pi\tau)=2$\,MHz. The seeded GSA solution $\phi(t)$ (blue solid) deviates significantly from the concatenated serrodyne seed $\phi_0(t)$ (red dashed). }
\label{fig:seedDivergence} 
\end{figure}

To gain some insight into how the differences between the concatenated serrodynes and GSA solutions affect a spectrum, we examine the phase modulation profiles in detail. The spectra of concatenated serrodynes contain unintended spectral content due to transients between the different serrodyne sections. For long waveform durations, these transients are sparse and hence the spectral deviations are small. The GSA solutions, when seeded with concatenated serrodyne solutions, then converge near the initial seeds (Fig.~\ref{fig:seedDivergence}a). However, for short waveform durations, switching transients occur often, and the GSA solutions significantly deviate from the initial concatenated serrodyne seeds. At waveform durations much shorter than $1/{\Delta \omega}$, where $\Delta \omega$ is the typical difference between two adjacent target frequencies, the GSA solution will no longer resemble the serrodyne seed but will still closely produce the target spectrum.

\section{Additional Considerations with Relevance to Optically Addressing Atoms and Molecules} \label{addConc}
While we have shown that multi-frequency spectra can be generated successfully (e.g. to address the $X(N=1) \rightarrow B(N=0)$ transition in CaF), producing the desired spectrum is not the only consideration when addressing optical transitions in atoms and molecules. While the energies of the internal states determine the required spectrum, the excited state lifetimes of atoms and molecules also affect the internal state dynamics and therefore provide additional timescales for the phase modulation profiles. For example, if one tries to address resolved hyperfine manifolds by concatenating a series of serrodynes, but at a rate much slower than the excited state decay rate, optical pumping into unaddressed manifolds will occur and molecules will fall into these dark states. This is very different from the case where all hyperfine frequency components are simultaneously present. This concern of falling into dark states is especially relevant in the context of photon cycling.

The question then becomes, given an excited state decay rate of $\Gamma$, what is the optimal waveform duration for the application at hand. In this section, we specifically consider the problem of maximizing the photon scattering rate, which is relevant to fluorescent detection and laser-cooling. Specifically, we will show that for the case of the $X(N=1) \rightarrow B(N=0)$ transition in CaF, GSA solutions and concatenated serrodynes can perform comparably and even better than simultaneous frequencies.

\subsection{Optical Bloch Equation Simulations}
We perform OBE simulations of the $X(N=1) - B(N=0)$ system in CaF in order to compare photon cycling rates obtained using different schemes of frequency generation. The goal in photon cycling is to maximize the photon scattering rate $\Gamma_{sc}$, which is related to the excited state population $P_e$ via $\Gamma_{sc} = \Gamma P_e$. For the $B$-state in CaF, $\Gamma=2\pi \times 6.3\,\text{MHz}$. In our OBE simulations, we calculate the time-averaged steady-state excited state population $P_e$ for concatenated serrodynes and GSA solutions with different waveform periods $\tau$ and light intensities $I/I_{\text{sat}}$, where $I_{\text{sat}} = \frac{\pi h c \Gamma}{3\lambda^3}$ is the saturation intensity. The results are compared to the case of simultaneous frequencies. 

To allow a fair comparison between the different schemes, the external fields and the light polarizations in the simulations are fixed. To avoid the formation of dark states, the following parameters are used. A small magnetic field of 5\,G is applied, and two beams of orthogonal linear polarizations are used. The polarizations are oriented 45 degrees with respect to the magnetic field, and the spectrum of the two beams are identical except for a 12\,MHz frequency shift. These parameters were found by optimizing for the case of simultaneous frequencies in the intensity range of $I/I_\text{sat} \approx 100-500$, where an excited state population of $P_e\approx0.19$ is reached. We note that under steady-state conditions, the theoretical maximum excited state population is given by $N_e/(N_g+N_e)$ where $N_e$ and $N_g$ are the number of states in the excited and ground manifolds, respectively. For the $X(N=1) \rightarrow B(N=0)$ transition, this ratio is 0.25. 

\begin{figure}[h!]
{\includegraphics[width=\columnwidth]{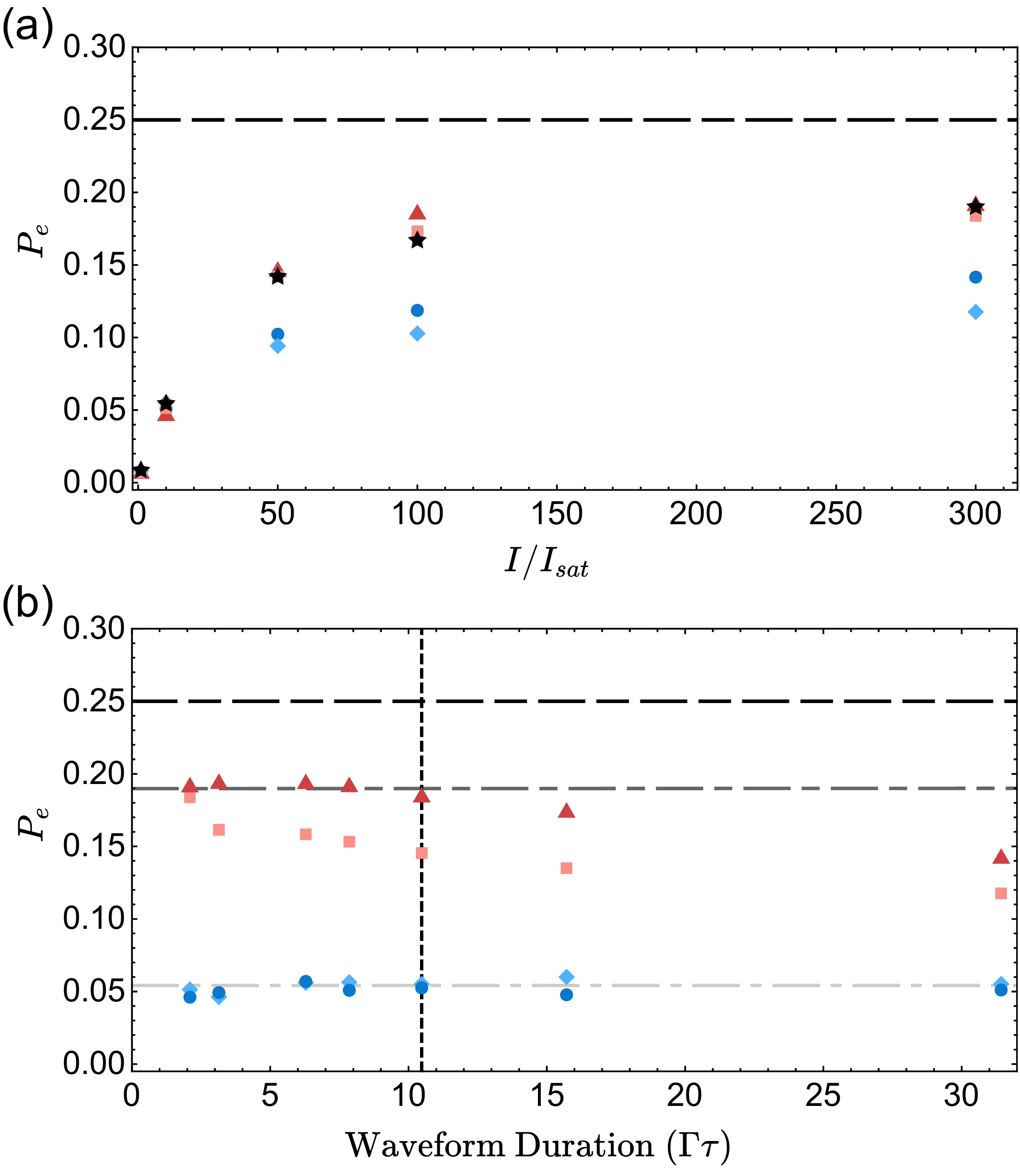}}
\caption{\label{fig:scatRate} Optical Bloch Equation Simulations for the $X(N=1) - B(N=0)$ System in CaF. (a) The average steady-state excited state population $P_e$ as a function of intensity $I/I_{\text{sat}}$. Shown in red triangles and light-red squares are the results of the GSA solution and the concatenated serrodynes, respectively, for a waveform duration of $\Gamma\tau=2.1$. For a longer waveform duration ($\Gamma\tau=31.4$), both the GSA solution (blue circles) and the concatenated serrodynes (light-blue diamonds) produce lower excited state populations. For comparison, black stars indicate the results for simultaneous frequencies and the black dashed line marks the theoretical maximum. (b) Excited state population $P_e$ as a function of waveform duration $\Gamma \tau$. At high intensity ($I/I_{\text{sat}}=300$), the GSA solutions (red triangles) shows higher $P_e$ than the concatenated serrodynes (light-red squares). At a lower intensity ($I/I_{\text{sat}}=10$), the GSA solutions (blue circles) and concatenated serrodynes (light-blue diamonds) yield similar $P_e$. For comparison, $P_e$ obtained using simultaneous frequencies at the corresponding high (dark-gray dashed line) and low intensities (light-gray dashed line) are shown. Also shown is the theoretical maximum (black dashed line). The vertical black dashed line denotes the waveform duration used in Fig.~\ref{fig:CaFSpec}.}
\end{figure}

Fig.~\ref{fig:scatRate} shows the excited population $P_e$ for simultaneous frequencies, concatenated serrodynes, and GSA solutions as a function of the intensity $I/I_{\text{sat}}$ and the dimensionless waveform duration $\Gamma\tau$. For short waveform durations, the excited state populations $P_e$ produced by concatenated serrodynes and GSA solutions are comparable to those produced by sending in simultaneous frequencies. At high intensities, the GSA solution and concatenated serrodynes both produce high excited state populations out to waveform durations of $\tau \approx 10 \Gamma^{-1}$, much longer than the simple expectation of $ \tau \approx \Gamma^{-1}$. In addition, we find that at high intensities ($I/I_{\text{sat}}=300$), GSA solutions produce higher excited state populations when compared to concatenated serrodynes. At lower intensities, the two methods produce similar results.

We believe that the long waveform duration threshold ($\tau \gg \Gamma^{-1}$) before the scattering rate decreases likely comes from the following three effects. First, the rate of going into an unaddressed hyperfine level is lower than $\Gamma$. When a single hyperfine manifold is addressed, the total population in the dark manifolds will grow at a rate of $\frac{N_{g}-N_A}{N_{g}} \Gamma$, where $N_{A}$ is the number of addressed states. The critical waveform duration ($\tau_c$) at which the dark state population becomes significant can be estimated by summing the inverses of these rates. For the present transition, this gives $\Gamma\tau_c=5.5$. Second, at $I/I_{\text{sat}}=300$, power broadening leads to broadened linewidths. Closely-spaced hyperfine manifolds, such as $F=2$ and $F=1^+$, are then effectively simultaneously addressed, which would further extend $\tau_c$. Third, the finite excited state lifetime implies that molecules experience a spectrum that is obtained from a time window of $\sim 1/\Gamma$ rather than the entire waveform. This means that the molecules see additional spectral content that can address multiple hyperfine manifolds simultaneously, further increasing $\tau_c$. In the case of photon scattering, the third effect likely leads to higher scattering rates in general, and could explain why the GSA solutions outperform the concatenated serrodynes at high intensities. 

To summarize, for the case of CaF at the aforementioned external fields and laser configurations, our simulations show that both concatenated serrodynes and GSA solutions produce excited state populations $P_e$ comparable to that obtained using simultaneous frequencies. We also find that these solutions continue to perform well at longer waveform durations ($\tau \gg \Gamma^{-1}$). In our setup, we have sufficient bandwidth to generate waveform durations as short as $ \tau=2 \Gamma^{-1}$ ($\tau=52\,\text{ns}$), implying that favorable scattering rates can be achieved. From the above considerations, we expect that for the purpose of cycling photons in molecules, our method is robust and will perform favorably for molecules with similar excited state lifetimes and ground state energy splittings.

\section{Summary and Outlook} \label{conc}
In this paper, we have shown that the problem of generating arbitrary optical spectra via phase modulation can be mapped to a computer-generated holography problem, which can be solved using existing methods such as GSA. We have further demonstrated the feasibility of this approach, and examined the additional considerations that arise when addressing atoms and molecules with phase-modulated light. We have verified through OBE simulations that photon cycling with spectra generated via GSA performs favorably compared to photon cycling with simultaneous frequencies.

In light of the present experimental push towards controlling molecules of ever-increasing complexity, we believe that the presented method is of practical value, since it simplifies the problem of generating $N$ arbitrary frequencies. Compared to existing solutions that require physical resources that scale with $N$, our method requires only a single device. Furthermore, the corresponding CGH problem can be solved efficiently using existing methods. Based on the features and limitations of our CGH-based spectral engineering method, we have identified the following scenarios where our method can be advantageous: 1) situations where a large number of arbitrary frequencies are needed, 2) situations where small spectral deviations at the $10^{-2}$ level can be tolerated, 3) situations where a broadband background is much more desirable than spurious peaks. The third situation is encountered when non-linearities are present. This occurs, for example, in laser slowing and atomic/molecular beam experiments, where the finite interaction time implies that spectral content below a certain intensity threshold is negligible. 

The situations we identified above already occur in ongoing searches for nuclear magnetic quadrupole moments (MQM) in molecules and in a recently proposed molecular slowing scheme. In a molecular MQM search, molecules with sufficient nuclear spin ($I\geq1$) are required, increasing the number of relevant optical transition frequencies~\cite{Flambaum1984EDM,Flambaum2014MQM,Kozyryev2017tviolation}. In a recently-proposed molecular Zeeman slower~\cite{Petzold2018}, multiple optical frequencies are also needed since large Zeeman splittings are encountered at the proposed magnetic fields. Beyond these two examples, our method could in the future open up molecular experiments that would otherwise be prohibitively complex.

%

\end{document}